\documentclass[aps,prd,twocolumn,10pt,superscriptaddress,nofootinbib,longbibliography]{revtex4-1}

\usepackage{amssymb}
\usepackage{graphicx}
\usepackage{amsmath}
\usepackage[colorlinks=true,linkcolor=royalblue,citecolor=magenta,urlcolor=magenta]{hyperref}
\usepackage{multirow}
\usepackage{setspace}
\usepackage{verbatim}
\usepackage{float}
\usepackage{color}
\usepackage{ulem}
\usepackage[utf8]{inputenc}
\ifdefined\DeclareUnicodeCharacter
  \DeclareUnicodeCharacter{2212}{\ensuremath{-}}
\fi
\usepackage[table,xcdraw]{xcolor}
\definecolor{royalblue}{rgb}{0.25,0.41,0.88}
\usepackage{makecell}
\usepackage{url}
\usepackage{bm}

\begin{document}
\title{Cosmological tests of chameleon dark energy with dark sector coupling}

\author{Jia-Qi Wang}
\email{wangjiaqi@itp.ac.cn}
\affiliation{Institute of Theoretical Physics, Chinese Academy of Sciences (CAS), Beijing 100190, China}
\affiliation{University of Chinese Academy of Sciences (UCAS), Beijing 100049, China}

\author{Zong-Kuan Guo}
\email{guozk@itp.ac.cn}
\affiliation{Institute of Theoretical Physics, Chinese Academy of Sciences (CAS), Beijing 100190, China}
\affiliation{University of Chinese Academy of Sciences (UCAS), Beijing 100049, China}
\affiliation{School of Fundamental Physics and Mathematical Sciences, Hangzhou Institute for Advanced Study, University of Chinese Academy of Sciences, Hangzhou 310024, China}

\author{Shao-Jiang Wang}
\email{schwang@itp.ac.cn}
\affiliation{Institute of Theoretical Physics, Chinese Academy of Sciences (CAS), Beijing 100190, China}
\affiliation{Asia Pacific Center for Theoretical Physics (APCTP), Pohang 37673, Korea}

\begin{abstract}
Planck cosmic microwave background (CMB) measurements and baryon acoustic oscillation (BAO) data from the Dark Energy Spectroscopic Instrument (DESI) give different estimates of the matter abundance within $\Lambda\mathrm{CDM}$.
Their combination with supernova data suggests an evolving dark energy equation of state that crosses $w=-1$, a behavior that cannot be realized by a single minimally coupled canonical scalar field.
This motivates testing the chameleon branch of nonminimally coupled quintessence, in which energy flows from dark energy to dark matter, using CMB, DESI DR2 BAO, and DES-Dovekie Type Ia supernova data.
The combined data favor a nonzero interaction at $3.3\sigma$ and provide comparable posterior support for the two directions of energy transfer under the adopted priors.
The chameleon branch improves the best-fit $\chi^2$ by $7.25$ relative to $\Lambda\mathrm{CDM}$.
Its modified dark matter evolution produces an effective dark energy equation of state that crosses from below to above $w=-1$ near $z\simeq0.5$ for the best-fit parameter combination, while the underlying scalar field remains canonical and never enters the phantom regime.
\end{abstract}


\keywords{chameleon dark energy, dark sector interactions, cosmological parameter constraints}
\maketitle

\section{Introduction}
\label{sec:introduction}

Observations of the cosmic microwave background (CMB)~\cite{WMAP:2012nax,Planck:2018vyg,ACT:2020gnv,AtacamaCosmologyTelescope:2025blo,AtacamaCosmologyTelescope:2025nti,SPT-3G:2025bzu}, baryon acoustic oscillations (BAO)~\cite{SDSS:2009ocz,BOSS:2016wmc,eBOSS:2020yzd,DESI:2024mwx,DESI:2025zgx}, and Type Ia supernovae (SN)~\cite{SDSS:2014iwm,Pan-STARRS1:2017jku,Scolnic:2021amr,Brout:2022vxf,Rubin:2023jdq,DES:2024jxu,DES:2025sig,Hoyt:2026fve} form the principal observational basis of the $\Lambda\mathrm{CDM}$ model.
As observational precision improves, differences emerge between the parameter values inferred from different probes within flat $\Lambda\mathrm{CDM}$.
BAO measurements from the Dark Energy Spectroscopic Instrument (DESI) Data Release 2 (DR2) favor a lower value of the present-day matter density parameter $\Omega_{\mathrm m}$ than that inferred from Planck CMB observations~\cite{DESI:2025zgx,Ye:2025ark}, and the Hubble constant $H_0$ inferred from Planck is lower than that measured by SH0ES using the distance ladder~\cite{Planck:2018vyg,Riess:2021jrx,Wang:2026jxd,Cai:2026swf}.
The dark energy (DE) equation of state (EoS) and density evolution have been constrained using a range of parametrizations~\cite{Chevallier:2000qy,Linder:2002et,Guo:2005ata,Guo:2006ab,Xia:2005ge,Zhao:2006bt,Xia:2006rr,Zhao:2006qg,Zhao:2007ew,Zhao:2009ti,Zhang:2017idq} and reconstructed using principal components, wavelets, correlated priors, and Gaussian processes~\cite{Crittenden:2005wj,Hojjati:2009ab,Crittenden:2011aa,Zhao:2012aw,Zhao:2017cud,Wang:2018fng,Keeley:2020aym,Raveri:2021dbu,Gu:2024jhl,Jiang:2024xnu,Liu:2026rot,Wang:2026sqy}.
Recent joint analyses based on these approaches favor evolution from $w<-1$ in the past to $w>-1$ today~\cite{Wang:2024qan,DESI:2025wyn,DESI:2025fii,Li:2025vuh}.
Although evidence for DE EoS crossing depends on the SN sample and calibration~\cite{Efstathiou:2024xcq,Huang:2025som,Popovic:2025glk,Wang:2026kor}, Planck CMB and DESI DR2 BAO alone show a roughly $2\sigma$ discrepancy within flat $\Lambda\mathrm{CDM}$, associated with different inferred values of $\Omega_{\mathrm m}$~\cite{DESI:2025zgx,Ye:2025ark}.
Crossing $w=-1$ was already discussed in early quintom models~\cite{Feng:2004ad,Feng:2004ff,Guo:2004fq,Zhao:2005vj,Guo:2006pc}, whereas the intrinsic EoS of a single minimally coupled canonical scalar field always satisfies $w\geq-1$~\cite{Vikman:2004dc}.
Comparisons indicate that models allowing such a crossing can bring the CMB and BAO inferences of the matter density into closer agreement than these single-field models can~\cite{DESI:2025fii,Shlivko:2026jxa}.

The connection between the inferred matter density and DE dynamics motivates investigating the evolution of the dark sector~\cite{Ye:2025ark,Li:2025eqh,Li:2025dwz,Yao:2025twv}.
In models with separately conserved cold dark matter (CDM), its density scales as $a^{-3}$, with the remaining dark sector density assigned to DE~\cite{Kunz:2007rk,Ye:2025ark}.
Nonminimal couplings between dark matter (DM) and DE can modify this scaling through energy exchange~\cite{Wetterich:1994bg,Amendola:1999er,Guo:2004vg,Guo:2004xx,Clemson:2011an,Wang:2016lxa,Li:2026asg}.
The resulting changes in the background expansion and structure growth can be constrained by CMB, BAO, SN, and large-scale structure (LSS) observations~\cite{Guo:2007zk,Wang:2014xca,Pettorino:2013oxa,Planck:2015bue,Gomez-Valent:2020mqn}.
Joint analyses incorporating DESI BAO now report a preference for these interactions~\cite{Giare:2024smz,Li:2024qso,Silva:2025hxw,Pan:2025qwy,Li:2025muv,Feng:2025mlo,Zhang:2025dwu,Samanta:2025oqz,Li:2026jgg}.
In the coupled models tested in Ref.~\cite{Li:2026xaz}, this preference is robust across the CMB data combinations and SN recalibrations considered.
Some models can also relax cosmological upper bounds on the neutrino mass~\cite{Wang:2025znm,Li:2026ldf} and improve agreement with measurements of the growth observable $f\sigma_8$~\cite{Wang:2025znm}.
Among these realizations, a conformal coupling between quintessence and DM can be derived from the matter action~\cite{Wetterich:1987fm,Damour:1994zq} and has been widely tested with Planck and subsequent cosmological data~\cite{Pettorino:2013oxa,Planck:2015bue,Gomez-Valent:2020mqn,Li:2023fdk}.

These indications of a nonzero interaction do not establish a unique direction of energy transfer across different models.
More general interaction histories allow the transfer direction to reverse at late times, as explored in reconstructions~\cite{Wang:2015wga,Yang:2015tzc,Li:2025ula} and extended coupling models~\cite{Cai:2009ht,Li:2025owk,Guedezounme:2025wav,Wang:2026wrk}.
For conformally coupled quintessence, recent joint analyses of CMB, DESI DR2 BAO, and SN observations favor a nonzero coupling with energy transfer from DM to DE at about $3\sigma$ in the marginalized posterior~\cite{Wang:2025znm,Li:2026xaz}.
In contrast, several studies have shown that energy transfer from DE to DM can produce crossing behavior in the effective DE EoS defined relative to standard $a^{-3}$ DM dilution, while the intrinsic EoS remains $w_\varphi\geq-1$~\cite{Das:2005yj,Chakraborty:2025syu}.
Moreover, the coupling that transfers energy from DE to DM can give rise to chameleon behavior with an inverse-power-law potential~\cite{Khoury:2003aq,Khoury:2003rn,Brax:2004qh,Boriero:2015loa}.
The effective mass then depends on the ambient density~\cite{Khoury:2013yya,Burrage:2017qrf}.
This environmental dependence has motivated proposals to alleviate the Hubble tension~\cite{Cai:2021wgv} and studies of the correlation between SN estimates of $H_0$ and the density contrast around their host galaxies~\cite{Yu:2022wvg}.
Related studies have also considered early dark energy (EDE)~\cite{Karwal:2016vyq,Poulin:2018cxd} triggered by a chameleon coupling~\cite{Karwal:2021vpk}, structure formation~\cite{Li:2009sy,Li:2010mqa}, and astrophysical and laboratory tests~\cite{Khoury:2013yya,Burrage:2017qrf}.
Unfortunately, the early-time effective mass can greatly exceed the Hubble parameter, which may cause severe numerical difficulties~\cite{Brax:2004qh,Boriero:2015loa}.
Previous analyses have used the adiabatic tracking approximation~\cite{Boriero:2015loa} or solved the full field evolution for different initial conditions~\cite{Chakraborty:2024xas,Gomez-Valent:2026ept}.
This motivates testing whether the chameleon branch is compatible with joint constraints on the background expansion and structure growth.

We focus on the chameleon branch of nonminimally coupled quintessence, denoted $\mathrm{NMCQ}^{+}$, in which energy flows from DE to DM.
We constrain the inverse-power-law potential index and coupling strength using recent cosmological observations, assuming that the field is initially at the minimum of the effective potential.
For comparison, we include the branch with energy transfer from DM to DE within the same framework, denoted $\mathrm{NMCQ}^{-}$~\cite{Wang:2025znm}, and compare their posterior support through joint sampling over both signs of the coupling.
We examine the inferred DM abundance and CMB power spectra, and interpret the effective DE evolution of the constrained models.

This paper is organized as follows.
Section~\ref{sec:model} presents the nonminimally coupled model and the chameleon mechanism.
Section~\ref{sec:methods} describes the numerical methods, data, and parameter inference.
Section~\ref{sec:results} presents the cosmological constraints and CMB spectra.
Section~\ref{sec:background-crossing} examines the background evolution and effective DE crossing.
We summarize the results and discuss their limitations and implications in Sec.~\ref{sec:conclusion}.
Throughout this paper, we use natural units with $c=\hbar=M_{\mathrm{Pl}}=1$, a metric signature $(-,+,+,+)$, and a scale factor normalized to $a_0=1$.

\section{The nonminimally coupled model}
\label{sec:model}

We consider a DE model in which a canonical scalar field is conformally coupled to DM~\cite{Boriero:2015loa,Chakraborty:2025syu}.
In the Einstein frame, the action is
\begin{equation}
\begin{split}
S={}&\int \mathrm{d}^4x\,\sqrt{-g}\left[
\frac{1}{2}R-\frac{1}{2}g^{\mu\nu}\partial_\mu\varphi\,\partial_\nu\varphi-V(\varphi)\right] \\
&+S_{\mathrm{DM}}\!\left[\psi_{\mathrm{DM}},\mathcal{A}^2(\varphi)g_{\mu\nu}\right]
+S_{\mathrm{SM}}\!\left[\psi_{\mathrm{SM}},g_{\mu\nu}\right],
\end{split}
\label{eq:action}
\end{equation}
where $S_{\mathrm{DM}}$ is the DM action and $S_{\mathrm{SM}}$ contains baryons, photons, and neutrinos, all minimally coupled to $g_{\mu\nu}$.
Only DM couples to $\widetilde g_{\mu\nu}=\mathcal{A}^2(\varphi)g_{\mu\nu}$.
We adopt the Ratra--Peebles inverse-power-law potential $V(\varphi)=V_*\varphi^{-n}$~\cite{Ratra:1987rm,Peebles:1987ek}, with $\varphi>0$, where $V_*>0$ sets the potential scale and $n>0$ is the potential index.
This inverse-power-law potential decreases monotonically with increasing field value.
We adopt the exponential conformal factor, $\mathcal{A}(\varphi)=e^{\beta \varphi}$~\cite{Wetterich:1987fm,Damour:1994zq}.

In a spatially flat FLRW background, the action in Eq.~\eqref{eq:action} gives the continuity equations for DM and the scalar field,
\begin{align}
\dot{\rho}_{\mathrm{DM}}+3H\rho_{\mathrm{DM}}&=\beta\dot\varphi\,\rho_{\mathrm{DM}},
\label{eq:cdm-eom}\\
\dot{\rho}_{\varphi}+3H(\rho_{\varphi}+p_\varphi)&=-\beta\dot\varphi\,\rho_{\mathrm{DM}},
\label{eq:scalar-eom}
\end{align}
where $\beta=\mathrm{d}\ln\mathcal{A}(\varphi)/\mathrm{d}\varphi$ denotes the constant, dimensionless coupling strength, and the sign of $\beta\dot\varphi$ determines the direction of energy transfer.
For trajectories with $\dot\varphi>0$, energy flows from the scalar field to DM when $\beta>0$, corresponding to $\mathrm{NMCQ}^{+}$, and from DM to the scalar field when $\beta<0$, corresponding to $\mathrm{NMCQ}^{-}$.
The scalar field has a canonical kinetic term and $V>0$, and its intrinsic EoS, $w_{\varphi}=p_\varphi/\rho_{\varphi}$, always satisfies $w_{\varphi}\geq-1$.

\begin{figure}[tp]
  \centering
  \includegraphics[width=\linewidth]{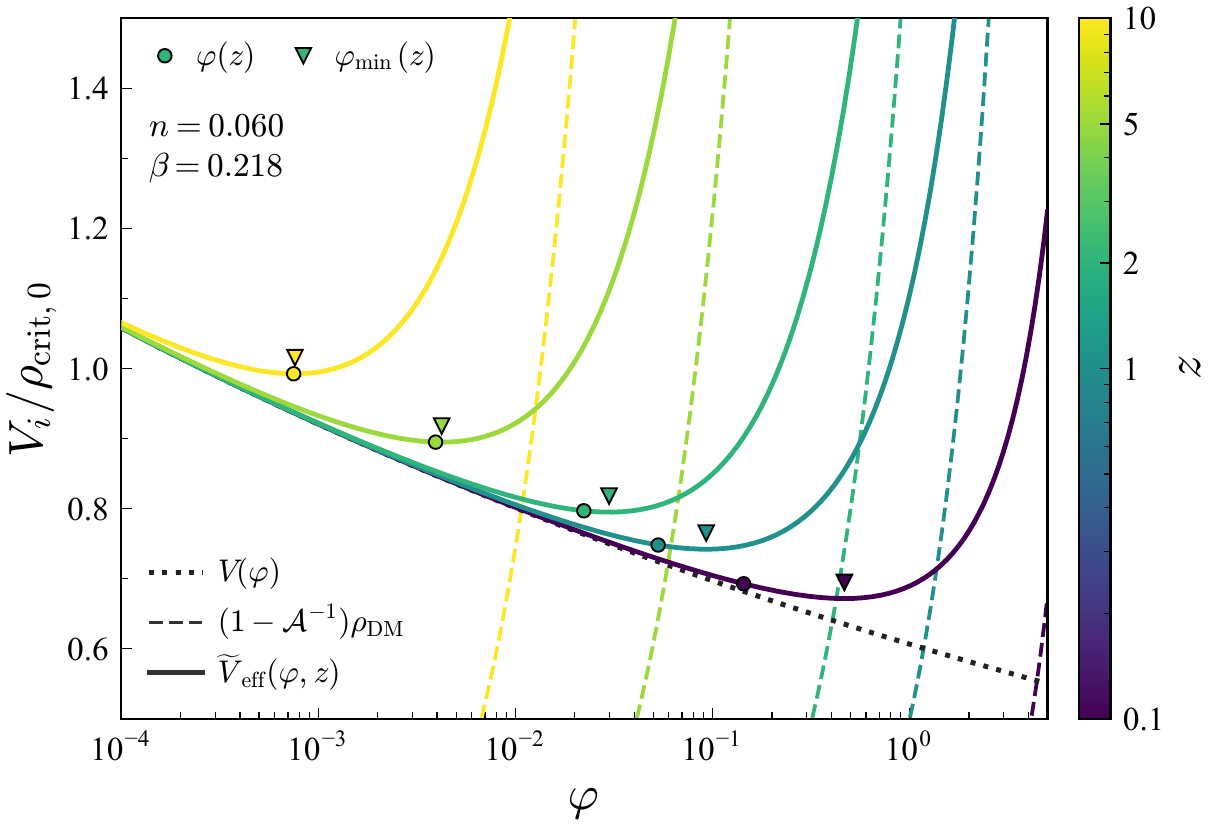}
  \caption{Effective potential of $\mathrm{NMCQ}^{+}$, evaluated at its best-fit point for the main data combination. The colors correspond to $z=0.1,1,2,5,10$. Dotted, dashed, and solid curves show $V$, $(1-\mathcal{A}^{-1})\rho_{\mathrm{DM}}$, and the shifted effective potential $\widetilde V_{\mathrm{eff}}=V+(1-\mathcal{A}^{-1})\rho_{\mathrm{DM}}$, respectively. All potential terms are normalized by $\rho_{\mathrm{crit},0}=3H_0^2$. Circles mark the actual field positions, and inverted triangles mark the corresponding instantaneous minima.}
  \label{fig:effective-potential}
\end{figure}

The coupling term in Eq.~\eqref{eq:scalar-eom} adds a term depending on the DM density to the inverse-power-law potential~\cite{Khoury:2003rn,Boriero:2015loa}, giving the effective potential $V_{\mathrm{eff}}(\varphi,a)=V(\varphi)+\rho_{\mathrm{DM}}(\varphi,a)$.
At fixed $a$, the field derivative satisfies $\partial_\varphi\rho_{\mathrm{DM}}=\beta\rho_{\mathrm{DM}}$.
For $\beta>0$, the inverse-power-law bare potential decreases with $\varphi$, whereas the DM contribution increases, yielding a unique instantaneous minimum,
\begin{equation}
 \varphi_{\min}(a)=\frac{n+1}{\beta}\,
 W\!\left[
 \frac{\beta}{n+1}
 \left(\frac{n V_* e^{\beta\varphi_0}}{\beta\rho_{\mathrm{DM},0}a^{-3}}\right)^{\!\frac{1}{n+1}}
 \right],
\label{eq:potential-minimum}
\end{equation}
where $W$ is the principal real branch of the Lambert function, satisfying $W(y)e^{W(y)}=y$~\cite{Cai:2021wgv}.
Here $\rho_{\mathrm{DM},0}\equiv\rho_{\mathrm{DM}}(a=1)$ and $\varphi_0\equiv\varphi(a=1)$ denote the present-day DM density and field value, respectively.
The effective mass is defined by the curvature of the effective potential,
\begin{equation}
\begin{split}
 m_{\mathrm{eff}}^2(\varphi,a)
 &\equiv\frac{\partial^2 V_{\mathrm{eff}}}{\partial\varphi^2}\\
 &=\frac{n(n+1)V(\varphi)}{\varphi^2}
 +\beta^2\rho_{\mathrm{DM}}(\varphi,a).
\end{split}
\label{eq:effective-mass}
\end{equation}
For $n>0$ and $\beta>0$, $m_{\mathrm{eff}}^2$ is strictly positive, ensuring stability at the minimum.
A higher ambient DM density leads to a smaller field value and a larger effective mass at the minimum.
This density dependence gives rise to chameleon behavior in the dark sector.
We denote the mass at the minimum by $m_{\mathrm{eff},\min}(a)\equiv m_{\mathrm{eff}}[\varphi_{\min}(a),a]$.
When $m_{\mathrm{eff},\min}\gg H$ and the minimum moves sufficiently slowly compared with the field response, the inverse-power-law bare potential gradient approximately balances the coupling term, allowing the field to adiabatically track the minimum.
As the hierarchy between the mass and the Hubble scale decreases, the field may depart from this balance, and $m_{\mathrm{eff}}$ evaluated along the actual trajectory then differs from $m_{\mathrm{eff},\min}$.
FIG.~\ref{fig:effective-potential} shows the effective potential of $\mathrm{NMCQ}^{+}$ at several redshifts.
As the Universe expands, the DM contribution decreases and the minimum shifts toward larger field values.

During slow evolution near the minimum, the large mass suppresses the quasistatic response of scalar fluctuations to density sources if $m_{\mathrm{eff}}\gg H$ and $k/a\ll m_{\mathrm{eff}}$.
The additional scalar force then has a range of approximately $m_{\mathrm{eff}}^{-1}$.
The density dependence of this interaction range underlies the local chameleon mechanism~\cite{Khoury:2003rn,Cai:2021wgv}.
In dense environments such as DM halos, the field may acquire a large effective mass, as in the early Universe.
A small inverse-power-law potential index can further suppress the fractional DE density perturbations, leaving DE smoother than DM~\cite{Boriero:2015loa}.

For $\beta<0$, both the inverse-power-law bare potential and the coupling term drive the field toward larger values.
The effective potential is therefore of the runaway type, with no minimum at finite $\varphi>0$.
The physical picture and cosmological constraints of $\mathrm{NMCQ}^{-}$ have been discussed in detail in Refs.~\cite{Pettorino:2013oxa,Planck:2015bue,Li:2023fdk,Wang:2025znm,Li:2026xaz}.

The conformal coupling also connects this background evolution to structure growth.
In addition to changing the DM density and expansion history, the coupling modifies the evolution of DM velocity perturbations, while spatial field fluctuations mediate an additional scalar interaction~\cite{Amendola:1999er,Boriero:2015loa,Uzan:2023dsk}.
The ratio $m_{\mathrm{eff}}/H$ characterizes the field response time relative to the background evolution, whereas the scalar force acting on a perturbation mode also depends on its physical wave number $k/a$ relative to $m_{\mathrm{eff}}$.
The background evolution and linear perturbations together determine the cosmological observables of the model.

\section{Methodology and data}
\label{sec:methods}

The nonminimal coupling makes the DM density depend on $\varphi$, so its normalization must be specified in the numerical calculation.
We adopt the convention that the input DM abundance corresponds to the physical density today, $\rho_{\mathrm{DM},0}$.
The corresponding parameter sampled in the analysis and reported in Tables~\ref{tab:priors} and~\ref{tab:main} is
\begin{equation}
\omega_{\mathrm{DM}}\equiv\Omega_{\mathrm{DM},0}h^2
=\frac{\rho_{\mathrm{DM},0}}{\rho_{\mathrm{crit},0}}h^2,
\label{eq:omega-dm}
\end{equation}
where $h=H_0/(100\,\mathrm{km\,s^{-1}\,Mpc^{-1}})$ and $\rho_{\mathrm{crit},0}=3H_0^2$. Eq.~\eqref{eq:cdm-eom} gives the normalized density evolution
\begin{equation}
\rho_{\mathrm{DM}}(a)=\rho_{\mathrm{DM},0}a^{-3}
e^{\beta[\varphi(a)-\varphi_0]}.
\label{eq:dm-normalization}
\end{equation}

Cosmological parameters are constrained using a modified version of CAMB\footnote{\url{https://github.com/cmbant/CAMB}}~\cite{Lewis:1999bs,Lewis:2026mif,Mead:2020vgs}, which includes energy and momentum transfer between DM and the scalar field in the background and perturbation equations~\cite{Li:2023fdk}.
To handle the strong numerical stiffness caused by the large early-time effective mass in $\mathrm{NMCQ}^{+}$, we implement a fourth-order, L-stable implicit Runge--Kutta method (ESDIRK).\footnote{\url{https://github.com/LLNL/sundials}}
The initial field value $\varphi_{\mathrm{ini}}$ is set at the minimum of the effective potential or on the attractor solution of $V(\varphi)$ in the early Universe ($a=10^{-12}$).
We use a shooting method to determine the inverse-power-law potential amplitude $V_*$ and present-day field value $\varphi_0$ to ensure consistency between the field equations and background densities.
Further details of the initial conditions and numerical methods are given in Appendices~\ref{sec:appendixA} and~\ref{sec:appendixB}.

We use Cobaya\footnote{\url{https://github.com/CobayaSampler/cobaya}}~\cite{Torrado:2020dgo} to perform nested sampling with PolyChord\footnote{\url{https://github.com/PolyChord/PolyChordLite}}~\cite{Handley:2015vkr} until the evidence remaining in the live points falls below $10^{-3}$ of the total evidence.
Table~\ref{tab:priors} lists the prior ranges of the cosmological parameters.
Posterior samples are analyzed using GetDist\footnote{\url{https://github.com/cmbant/getdist}}~\cite{Lewis:2019xzd}.
The best-fit parameters are those that yield the lowest total $\chi^2$.
The difference in best-fit $\chi^2$ relative to $\Lambda$CDM is $\Delta\chi^2_{\min}=\chi^2_{\min}(\mathrm{model})-\chi^2_{\min}(\Lambda\mathrm{CDM})$.

\begin{table}[bt]
\centering
\caption{Uniform prior ranges for the cosmological parameters.}
\label{tab:priors}
\begin{minipage}{0.60\columnwidth}
\begin{ruledtabular}
\begin{tabular}{lc}
Parameter & Prior range \\
\hline
$100\omega_{\mathrm b}$ & $[2.0,2.4]$ \\
$100\omega_{\mathrm{DM}}$ & $[10,13]$ \\
$100h$ & $[60,75]$ \\
$\tau$ & $[0.02,0.20]$ \\
$\ln(10^{10}A_s)$ & $[2.9,3.2]$ \\
$n_s$ & $[0.93,1.00]$ \\
\hline
$n$ & $[0.01,2]$ \\
$\beta$ & $[-0.25,0.4]$ \\
\end{tabular}
\end{ruledtabular}
\end{minipage}
\end{table}

The datasets used in this analysis are summarized below:
\begin{itemize}
\item CMB: CamSpec high-multipole TT, TE, and EE spectra based on Planck PR4 NPIPE~\cite{Rosenberg:2022sdy}, together with the Planck 2018 low-multipole temperature and $E$-mode polarization likelihoods~\cite{Planck:2019nip} and the joint Planck PR4 and ACT DR6 CMB lensing reconstruction~\cite{Carron:2022eyg,ACT:2023kun}.
\item BAO: DESI DR2 BAO distance measurements and their covariance matrix~\cite{DESI:2025zgx} are used to constrain distance scales relative to the sound horizon at the baryon drag epoch.
\item SN Ia: DES-Dovekie supernova sample~\cite{DES:2025sig}, including its statistical and systematic covariance.
\end{itemize}
These datasets form the baseline combination. Appendix~\ref{sec:appendixC} presents results with Pantheon+~\cite{Brout:2022vxf} or Union3.1~\cite{Hoyt:2026fve} replacing DES-Dovekie to assess the dependence on the SN sample.

\section{Cosmological constraints}
\label{sec:results}
\label{sec:constraints}

\begin{figure}[tp]
  \centering
  \includegraphics[width=\linewidth]{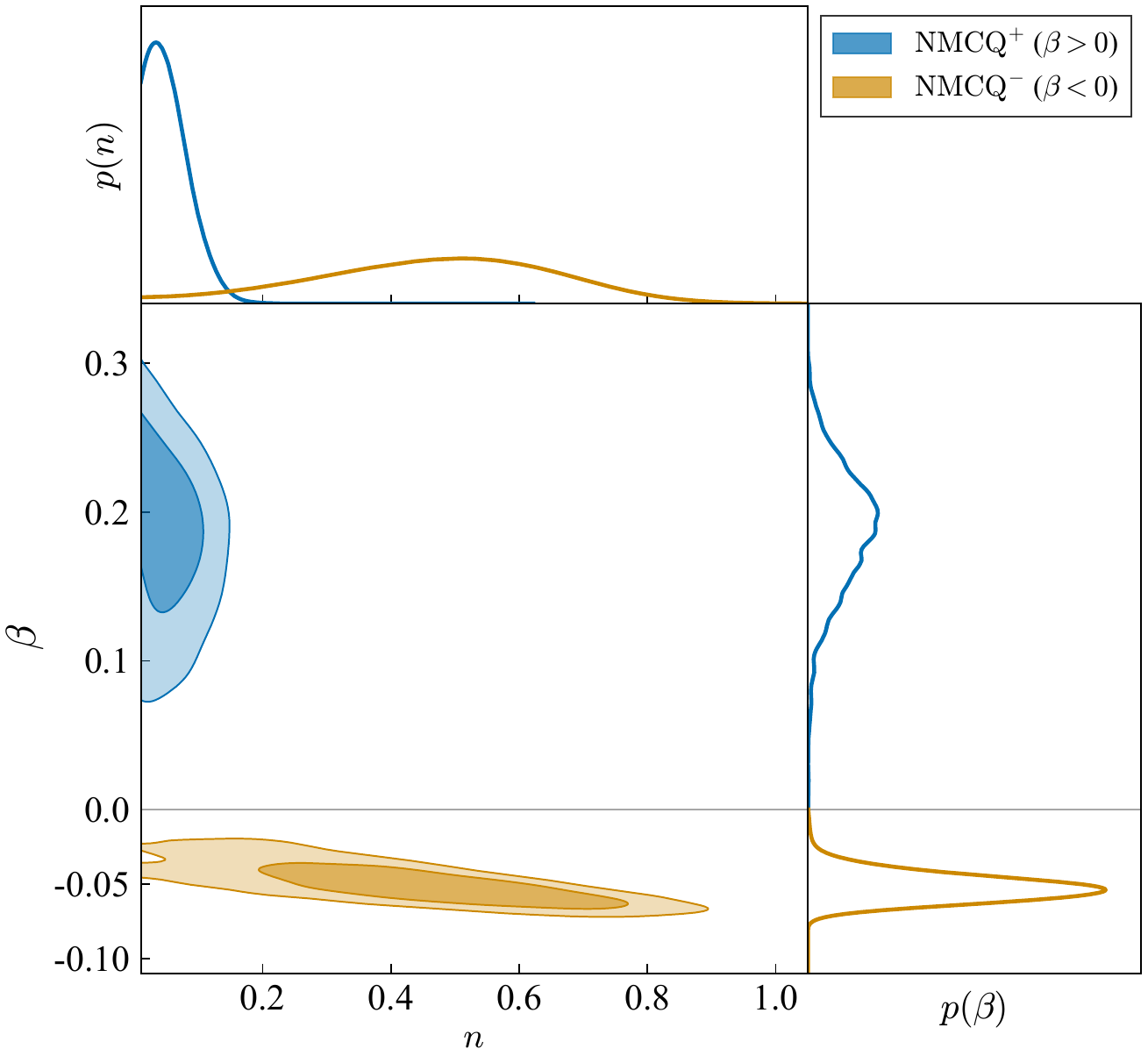}
  \caption{Constraints on the inverse-power-law potential index $n$ and coupling strength $\beta$ from the main data combination. Blue and orange denote $\mathrm{NMCQ}^{+}$ and $\mathrm{NMCQ}^{-}$, respectively, and the horizontal gray line marks $\beta=0$. Dark and light shaded regions enclose 68\% and 95\% of the conditional posterior probability. The two-dimensional distributions are obtained from separate sampling for each branch and are normalized separately. The one-dimensional distributions are obtained from joint sampling over both signs of the coupling and share a common normalization across the two branches.}
  \label{fig:alpha-beta}
\end{figure}

The baseline datasets favor a small inverse-power-law potential index, $n\lesssim0.1$, and a nonzero coupling $\beta$ at approximately $3\sigma$ in the marginalized posterior for the chameleon branch, as summarized in Table~\ref{tab:main}.
For comparison, $\mathrm{NMCQ}^{-}$ allows a larger inverse-power-law potential index but a weaker coupling.
Both branches improve the best-fit total $\chi^2$ relative to $\Lambda\mathrm{CDM}$, with $\Delta\chi^2_{\min}=-7.25$ for $\mathrm{NMCQ}^{+}$ and $-7.84$ for $\mathrm{NMCQ}^{-}$.
The difference between the branches is only $-0.59$ in favor of $\mathrm{NMCQ}^{-}$.
Relative to $\mathrm{NMCQ}^{+}$, the lower BAO and SN contributions to $\chi^2$ in $\mathrm{NMCQ}^{-}$ are partly offset by its higher CMB contribution, including lensing.
The two coupling directions also receive similar integrated posterior support in the joint analysis, as shown in FIG.~\ref{fig:alpha-beta}.
With a uniform prior $\beta\in[-0.25,0.4]$ covering the $3\sigma$ posterior ranges of both branches, the posterior probabilities of $\mathrm{NMCQ}^{+}$ and $\mathrm{NMCQ}^{-}$ are $49.1\%$ and $50.9\%$, respectively.
The joint marginalized $\beta$ posterior favors a nonzero coupling at a Gaussian-equivalent level of $3.3\sigma$.
The full cosmological parameter posteriors are presented in Appendix~\ref{sec:appendixC}.

\begin{table}[b]
\raggedright
\let\flushing\raggedright
\caption{Cosmological parameter constraints from the baseline datasets. Para. denotes parameter. Results are quoted as posterior means with 68\% credible intervals, or as 68\% upper limits. $S_8$ is defined as $S_8=\sigma_8(\Omega_{\mathrm m}/0.3)^{1/2}$.}
\label{tab:main}
\label{tab:cosmological_constraints}
\renewcommand{\arraystretch}{1.25}
{\small
\setlength{\tabcolsep}{2pt}
\begin{tabular*}{\columnwidth}{@{\extracolsep{\fill}}llll@{}}
\hline
Para. & $\Lambda$CDM & \begin{tabular}[c]{@{}l@{}}$\mathrm{NMCQ}^{+}$\\$(\beta>0)$\end{tabular} & \begin{tabular}[c]{@{}l@{}}$\mathrm{NMCQ}^{-}$\\$(\beta<0)$\end{tabular} \\
\hline
$100\omega_{\mathrm b}$ & $2.233\pm 0.012$ & $2.246\pm 0.013$ & $2.218\pm 0.014$ \\
$100\omega_{\mathrm{DM}}$ & $11.789\pm 0.060$ & $11.98\pm 0.10$ & $11.51\pm 0.13$ \\
$100h$ & $68.07\pm 0.27$ & $67.57^{+0.50}_{-0.41}$ & $67.86\pm 0.56$ \\
$n_{\mathrm s}$ & $0.9686\pm 0.0033$ & $0.9722\pm 0.0036$ & $0.9655\pm 0.0038$ \\
$n$ & $\cdots$ & $<0.0703$ & $0.44\pm 0.18$ \\
$\beta$ & $\cdots$ & $0.186^{+0.047}_{-0.038}$ & $-0.0504^{+0.0078}_{-0.012}$ \\
\hline
$\Omega_{\mathrm m}$ & $0.3040\pm 0.0035$ & $0.3132^{+0.0055}_{-0.0067}$ & $0.2995\pm 0.0053$ \\
$S_8$ & $0.8142\pm 0.0072$ & $0.873\pm 0.025$ & $0.8220\pm 0.0084$ \\
\hline
$\Delta\chi^2_{\min}$ & $0$ & $-7.25$ & $-7.84$ \\
\hline
\end{tabular*}
}
\end{table}

The coupling direction directly affects the inferred present-day DM density.
Table~\ref{tab:main} shows that $\rho_{\mathrm{DM},0}$ shifts upward in $\mathrm{NMCQ}^{+}$ and downward in $\mathrm{NMCQ}^{-}$ relative to $\Lambda\mathrm{CDM}$.
These shifts arise because the background evolution around recombination is largely fixed by the CMB.
Under the assumption of distance duality, BAO and SN observations also tightly constrain the low-redshift distance--redshift relation up to an overall distance calibration~\cite{Tiwari:2026pzk,Ling:2025lmw}.
The present-day DM normalization must therefore accommodate the modified dilution in Eq.~\eqref{eq:dm-normalization}~\cite{Boriero:2015loa}.

\begin{figure}[tp]
  \centering
  \includegraphics[width=\linewidth]{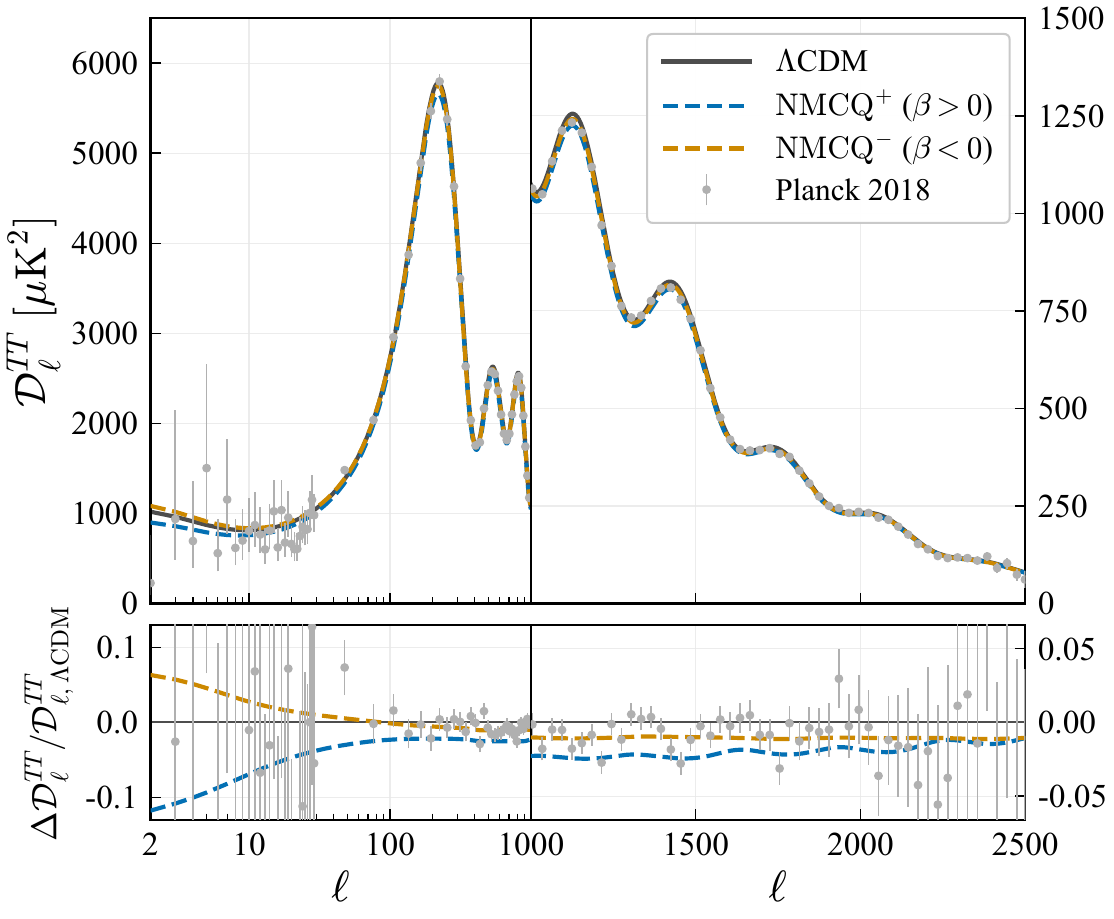}
  \caption{CMB temperature spectra for the best-fit parameter combinations of $\mathrm{NMCQ}^{+}$, $\mathrm{NMCQ}^{-}$, and $\Lambda\mathrm{CDM}$ for the main data combination, over $2\leq\ell\leq2500$. Upper panels: lensed $\mathcal D_\ell^{TT}\equiv\ell(\ell+1)C_\ell^{TT}/(2\pi)$ in $\mu\mathrm K^2$. Lower panels: dimensionless deviations $\Delta\mathcal D_\ell/\mathcal D_{\ell,\Lambda\mathrm{CDM}}$ of the two coupled models and the observational reference points from the same $\Lambda\mathrm{CDM}$ spectrum. Gray points show the public Planck 2018 TT spectrum~\cite{Planck:2019nip}. The left and right columns use logarithmic and linear horizontal scales, respectively, with different vertical ranges.}
  \label{fig:cmb-spectrum}
\end{figure}

FIG.~\ref{fig:cmb-spectrum} shows the effects of the nonminimally coupled models on the CMB temperature spectra.
The acoustic peak positions remain broadly similar in all three models, while the TT power in $\mathrm{NMCQ}^{+}$ is about $2.2\%$--$2.5\%$ lower than in $\Lambda\mathrm{CDM}$ over $200\lesssim\ell\lesssim1500$.
For the best-fit parameter combination, $A_s e^{-2\tau}$ is about $2.7\%$ lower than in $\Lambda\mathrm{CDM}$. This reduction is comparable to that in the acoustic-scale power, whereas the larger $n_s$ increases the relative power on small scales.
By comparison, the $\mathrm{NMCQ}^{-}$ spectrum shows smaller deviations over most acoustic scales.
On the largest angular scales, TT power is lower in $\mathrm{NMCQ}^{+}$ and higher in $\mathrm{NMCQ}^{-}$.
These scales are also sensitive to the late-time integrated Sachs--Wolfe (ISW) contribution from evolving gravitational potentials, which respond differently to the two directions of energy transfer~\cite{Amendola:1999er,Boriero:2015loa}.

Neither coupling direction substantially changes the inferred $H_0$ relative to $\Lambda\mathrm{CDM}$, as shown in Table~\ref{tab:main}.
For the adopted initial conditions, slow field evolution before recombination gives $|\beta\dot\varphi|\ll H$, so the interaction term in Eq.~\eqref{eq:cdm-eom} is small compared with Hubble dilution.
Together with the CMB constraint on the early matter content, this keeps the sound horizon $r_s$ close to its $\Lambda\mathrm{CDM}$ value.
Late-time expansion is further restricted by the BAO and SN distance constraints~\cite{Tiwari:2026pzk,Cai:2021weh}.
The larger $S_8$ of $\mathrm{NMCQ}^{+}$ does not by itself establish a conflict with weak-lensing constraints inferred assuming $\Lambda\mathrm{CDM}$~\cite{Wright:2025xka}.
As a chameleon candidate, $\mathrm{NMCQ}^{+}$ may produce observable local signatures through density-dependent changes in the scalar force and DM structure~\cite{Li:2010mqa}.

\section{Background evolution and effective crossing}
\label{sec:background-crossing}

\begin{figure}[tp]
  \centering
  \includegraphics[width=\linewidth]{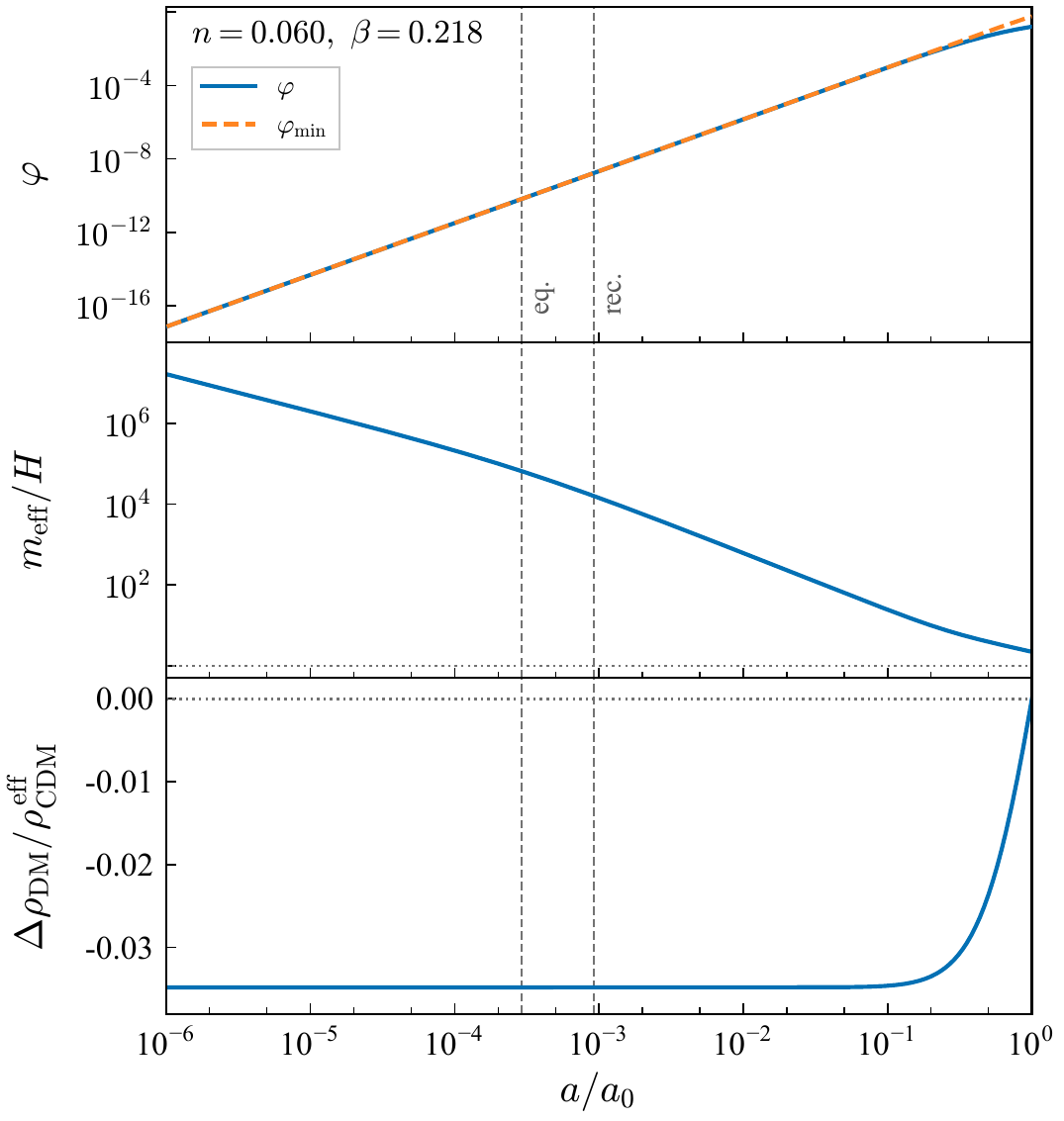}
  \caption{Background evolution of $\mathrm{NMCQ}^{+}$. Top: actual field trajectory $\varphi$ (solid) and instantaneous minimum $\varphi_{\min}$ (dashed). Middle: $m_{\mathrm{eff}}/H$ evaluated along the actual trajectory, with the horizontal dotted line marking $m_{\mathrm{eff}}/H=1$. Bottom: the fractional DM density difference $\Delta\rho_{\mathrm{DM}}/\rho^{\mathrm{eff}}_{\mathrm{CDM}}$, with the horizontal dotted line marking the present-day value $\Delta\rho_{\mathrm{DM}}=0$.}
  \label{fig:field-evolution}
\end{figure}

In this section, we focus on the background evolution of the chameleon branch, $\mathrm{NMCQ}^{+}$.
With the best-fit parameter combination, the scalar field evolves as shown in FIG.~\ref{fig:field-evolution}.
The field closely tracks the effective potential minimum until $z\sim10$.
In this regime, Eq.~\eqref{eq:potential-minimum} gives $\mathrm{d}\ln V/\mathrm{d}\ln a\simeq-3n/(1+n+\beta\varphi)$. Therefore, the small value of $n$ corresponds to slowly varying potential energy along the tracking solution.
At late times, $m_{\mathrm{eff}}/H$ falls to $\mathcal{O}(1)$ and the trajectory departs from the moving minimum.
The scalar effective mass satisfies $m_{\mathrm{eff}}>H$ throughout the evolution shown, with $m_{\mathrm{eff}}/H\simeq2.23$ at $z=0$.

Nonminimal coupling modifies the DM dilution law, making the separation of DM and DE in the interpretation of observations nontrivial.
To assess whether $\mathrm{NMCQ}^{+}$ can reproduce the DE evolution inferred from observations within minimally coupled models such as $w_0w_a\mathrm{CDM}$, we separate an effective CDM component from the total dark sector density~\cite{Das:2005yj,Boriero:2015loa,Wang:2026wrk}.
The reference density is defined as
\begin{equation}
\rho^{\mathrm{eff}}_{\mathrm{CDM}}(a)
\equiv\rho_{\mathrm{DM},0}a^{-3},
\label{eq:effective-cdm}
\end{equation}
which equals the physical DM density at the present epoch but generally differs from it at earlier times.
The departure of the physical density from this reference is denoted by $\Delta\rho_{\mathrm{DM}}\equiv\rho_{\mathrm{DM}}-\rho^{\mathrm{eff}}_{\mathrm{CDM}}$.
Eq.~\eqref{eq:dm-normalization} gives the fractional difference
\begin{equation}
\frac{\Delta\rho_{\mathrm{DM}}}{\rho^{\mathrm{eff}}_{\mathrm{CDM}}}
=e^{\beta(\varphi-\varphi_0)}-1.
\label{eq:dm-density-difference}
\end{equation}
With $\beta>0$ and a monotonically increasing field, $\varphi<\varphi_0$ at earlier times, and this definition gives $\Delta\rho_{\mathrm{DM}}<0$.
FIG.~\ref{fig:field-evolution} shows the evolution of the fractional DM density difference.
In the early Universe, including recombination, the fractional deficit relative to $\rho^{\mathrm{eff}}_{\mathrm{CDM}}$ remains nearly constant at about $3.5\%$.
At the background level, $\Delta\rho_{\mathrm{DM}}$ therefore behaves approximately as a negative CDM contribution scaling as $a^{-3}$.
Its leading effect can be compensated by the fitted DM abundance~\cite{Boriero:2015loa}, consistent with the similar CMB acoustic peak positions in FIG.~\ref{fig:cmb-spectrum}.
At late times, the DM density dilutes more slowly than $a^{-3}$, while its negative difference from the reference increases toward zero, giving $(\Delta\rho_{\mathrm{DM}})^{\cdot}>0$.
This positive contribution can offset the decline in $\rho_\varphi$ and allow an effective crossing of $w=-1$.

Subtracting the reference density from the total dark sector density defines the effective DE density, with its EoS given by the continuity equations,
\begin{equation}
\begin{split}
\rho_{\mathrm{DE}}^{\mathrm{eff}}
&\equiv\rho_\varphi+\Delta\rho_{\mathrm{DM}},\\
w_{\mathrm{DE}}^{\mathrm{eff}}
&\equiv-1-\frac{\dot{\rho}^{\mathrm{eff}}_{\mathrm{DE}}}{3H\rho_{\mathrm{DE}}^{\mathrm{eff}}}
=-1+\frac{\dot\varphi^2+\Delta\rho_{\mathrm{DM}}}{\rho_{\mathrm{DE}}^{\mathrm{eff}}}.
\end{split}
\label{eq:effective-de}
\end{equation}
This decomposition preserves the total dark sector density and its contribution to the expansion rate.
For $\rho_{\mathrm{DE}}^{\mathrm{eff}}>0$, Eq.~\eqref{eq:effective-de} gives $w_{\mathrm{DE}}^{\mathrm{eff}}<-1$ when this deficit satisfies $\Delta\rho_{\mathrm{DM}}<-\dot\varphi^2$, while the intrinsic scalar field EoS remains $w_\varphi\geq-1$.

\begin{figure}[tp]
  \centering
  \includegraphics[width=\linewidth]{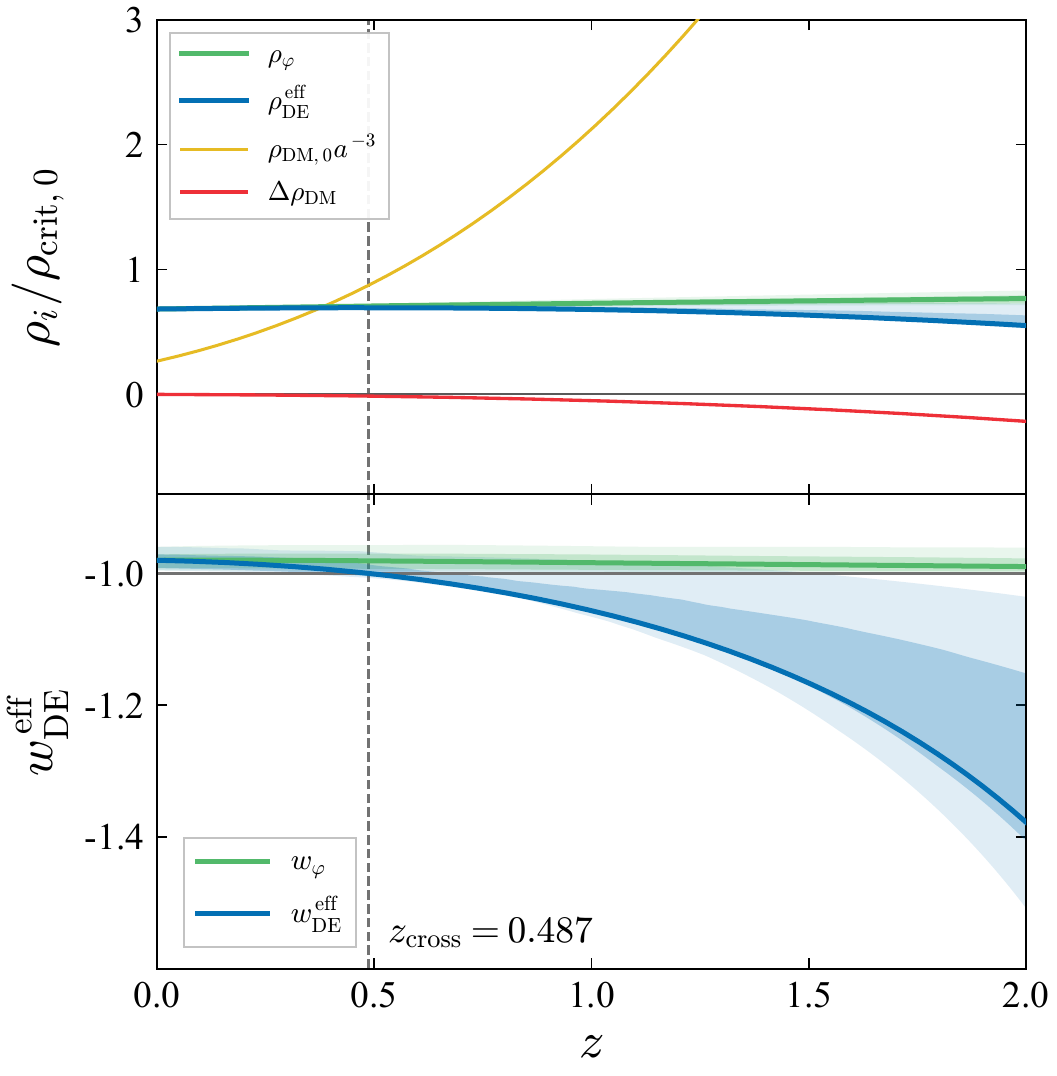}
  \caption{Evolution of the densities and EoS parameters in $\mathrm{NMCQ}^{+}$. The upper panel shows the scalar field density $\rho_\varphi$, the effective density $\rho_{\mathrm{DE}}^{\mathrm{eff}}$, the reference density $\rho^{\mathrm{eff}}_{\mathrm{CDM}}=\rho_{\mathrm{DM},0}a^{-3}$, and the DM density difference $\Delta\rho_{\mathrm{DM}}=\rho_{\mathrm{DM}}-\rho^{\mathrm{eff}}_{\mathrm{CDM}}$, all normalized to $\rho_{\mathrm{crit},0}$. The lower panel shows $w_\varphi$ and $w_{\mathrm{DE}}^{\mathrm{eff}}$. Solid curves show the best-fit solution, and dark and light bands indicate the 68\% and 95\% pointwise posterior intervals, respectively.}
  \label{fig:rho-w}
\end{figure}

FIG.~\ref{fig:rho-w} shows the effective DE EoS parameter for the best-fit $\mathrm{NMCQ}^{+}$ parameter combination.
The effective DE EoS crosses the quintom divide, $w=-1$, from below to above at $z_{\mathrm{cross}}\simeq0.487$.
This result is in close agreement with the crossing near $z\simeq0.5$ inferred from DESI DR2 BAO combined with CMB and SN data~\cite{DESI:2025fii}.
Note that this effective DE is defined only to compare observables between models and does not correspond to a physical component in the action, Eq.~\eqref{eq:action}.
Its crossing behavior therefore does not imply that any physical component increases in density as the Universe expands.
Furthermore, in the earlier DM-dominated regime, $\rho_\varphi=-\Delta\rho_{\mathrm{DM}}$ at $z\simeq3.52$, giving $\rho_{\mathrm{DE}}^{\mathrm{eff}}=0$.
The resulting divergence in $w_{\mathrm{DE}}^{\mathrm{eff}}$ reflects this definition, while the physical densities and expansion rate remain regular.

Comparisons between models must also account for their inferred DM abundance and total dark sector density.
In $\mathrm{NMCQ}^{-}$, the lower $\omega_{\mathrm{DM}}$ can compensate for differences in DE evolution, allowing an expansion history close to that associated with crossing in $w_0w_a\mathrm{CDM}$ even when $w_{\mathrm{DE}}^{\mathrm{eff}}$ remains above $-1$~\cite{Wang:2025znm}.
However, for the baseline datasets, $\omega_{\mathrm{DM}}$ in $\mathrm{NMCQ}^{+}$ is close to the $w_0w_a\mathrm{CDM}$ result in Ref.~\cite{Wang:2026wrk}.
This provides an approximate basis for comparing their low-redshift effective DE evolution.

\section{Conclusion and discussion}
\label{sec:conclusion}
\label{sec:discussion}

The chameleon branch of nonminimally coupled quintessence, $\mathrm{NMCQ}^{+}$, is constrained in this work using CMB, DESI DR2 BAO, and DES-Dovekie SN data, which favor a small inverse-power-law potential index, $n<0.0703$, and a nonzero coupling, $\beta=0.186^{+0.047}_{-0.038}$, at the 68\% credible level.
The data yield a best-fit improvement of $\Delta\chi^2_{\min}=-7.25$ for $\mathrm{NMCQ}^{+}$ relative to $\Lambda\mathrm{CDM}$, close to $-7.84$ for the opposite coupling branch, $\mathrm{NMCQ}^{-}$.
A nonzero coupling is favored at a Gaussian-equivalent level of $3.3\sigma$ in the joint marginalized posterior under the adopted model and priors.
The modified DM dilution in $\mathrm{NMCQ}^{+}$ produces an effective DE EoS that crosses $w=-1$ near $z\simeq0.5$, while the intrinsic scalar field EoS remains $w_\varphi\geq-1$.
The inferred $H_0$ remains close to the $\Lambda\mathrm{CDM}$ value, while $S_8$ is higher.

The choice of initial conditions can substantially affect the field evolution.
In this work, the field is initialized at the minimum of the effective potential.
The inverse-power-law potential and nonminimal coupling then jointly govern the field evolution during an extended tracking phase, before the late-time departure shown in FIG.~\ref{fig:field-evolution}.
Other initial conditions can produce delayed field evolution or reversals, as discussed for chameleon models~\cite{Chakraborty:2024xas,Gomez-Valent:2026ept} and related models with exponential potentials~\cite{Bedroya:2025fwh}.
Constraining these alternative histories requires a separate fit to the data.

As an important example of chameleon DE, $\mathrm{NMCQ}^{+}$ may exhibit environmental effects beyond the homogeneous background and linear perturbations.
A large scalar mass in dense environments can suppress the contribution of the interior of a sufficiently large body to its external field, reducing the fifth force even for matter couplings of gravitational strength~\cite{Khoury:2003rn}.
Related couplings, which link the scalar dynamics to the onset of matter domination, have also been explored in chameleon early DE~\cite{Karwal:2021vpk}.
Proposed effects of chameleon environments on local expansion~\cite{Cai:2021wgv}, together with simulations of halo structure and voids in related models~\cite{Li:2010mqa,Li:2010zzx}, motivate further tests using clustering and weak lensing.
Assessing screening and these environmental effects within the allowed $\mathrm{NMCQ}^{+}$ parameter region requires solutions for the inhomogeneous scalar field and matter distribution.
Investigating these effects and their possible impact on local determinations of $H_0$ is a direction for future research.

\begin{acknowledgments}
We are grateful to Wang-Wei Yu, Zi-Yan Yuwen and Jia-Le Ling for insightful discussions. 
This work is supported by the National Key Research and Development Program of China Grants (No. 2021YFA0718304, No. 2021YFC2203004, and No. 2020YFC2201501), 
the National Natural Science Foundation of China Grants (No. 12422502, No. 12547110, No. 12588101, No. 12475046, No. 12447101, No. 12475067 and No. 12235019),
and the China Manned Space Program Grant No. CMS-CSST-2025-A01 and No. CMS-CSST-2025-A02.
We also acknowledge the use of the HPC Cluster of ITP-CAS.
\end{acknowledgments}

\appendix

\section{Initial conditions and field evolution}
\label{sec:appendixA}

Initial conditions are set at $a_{\mathrm{ini}}=10^{-12}$ during radiation domination.
Two choices are considered.
One choice is set by the minimum of the effective potential.
For $\beta>0$, its position is
\begin{equation}
 \varphi_{\min}(a)=\frac{n+1}{\beta}\,
 W\!\left[
 \frac{\beta}{n+1}
 \left(\frac{n V_* a^3 e^{\beta \varphi_0}}{\beta\rho_{\mathrm{DM},0}}\right)^{\!\frac{1}{n+1}}
 \right].
 \label{eq:appendix-minimum-lambert}
\end{equation}
Here $W(y)$ is the principal real branch of the Lambert $W$ function, defined by $W(y)e^{W(y)}=y$.
Furthermore,
\begin{equation*}
 W'(y)\equiv\frac{\mathrm dW}{\mathrm dy}
 =\frac{W(y)}{y[1+W(y)]},
\end{equation*}
which gives
\begin{equation}
 \begin{aligned}
  \frac{\mathrm d\varphi_{\min}}{\mathrm d\ln a}
  &=\frac{3\varphi_{\min}}{n+1+\beta\varphi_{\min}},\\
  \varphi_{\min}'&=\mathcal H\frac{3\varphi_{\min}}{n+1+\beta\varphi_{\min}},
 \end{aligned}
 \label{eq:appendix-minimum-initial}
\end{equation}
where $\mathcal H=aH$, and primes on dynamical variables denote derivatives with respect to conformal time $\eta$.

The other choice is the attractor solution for the inverse-power-law bare potential during radiation domination~\cite{Ratra:1987rm,Peebles:1987ek},
\begin{equation}
 \begin{aligned}
  \varphi_{\rm attr}(a)&=
  \left[\frac{n(n+2)^2V_*}{4H^2(a)(n+6)}\right]^{1/(n+2)},\\
  \varphi_{\rm attr}'(a)&=\frac{4aH(a)}{n+2}\varphi_{\rm attr}(a).
 \end{aligned}
 \label{eq:appendix-attractor-initial}
\end{equation}
Both $\varphi_{\min}$ and $\varphi_{\rm attr}$ exist for $\mathrm{NMCQ}^{+}$, and the smaller value at the initial time is chosen,
\begin{equation}
 \varphi_{\mathrm{ini}}=
 \min\!\left[\varphi_{\min}(a_{\mathrm{ini}}),
 \varphi_{\rm attr}(a_{\mathrm{ini}})\right].
 \label{eq:appendix-initial-choice}
\end{equation}
Since $V\propto\varphi^{-n}$ decreases with $\varphi$, this choice gives the larger inverse-power-law bare potential energy of the two candidates.
For $\mathrm{NMCQ}^{-}$, the initial condition $\varphi_{\mathrm{ini}}=\varphi_{\rm attr}(a_{\mathrm{ini}})$ is adopted.
The initial field velocity in both models is the conformal time derivative of the selected solution at $a_{\mathrm{ini}}$.

\begin{figure}[t]
 \centering
 \includegraphics[width=\columnwidth]{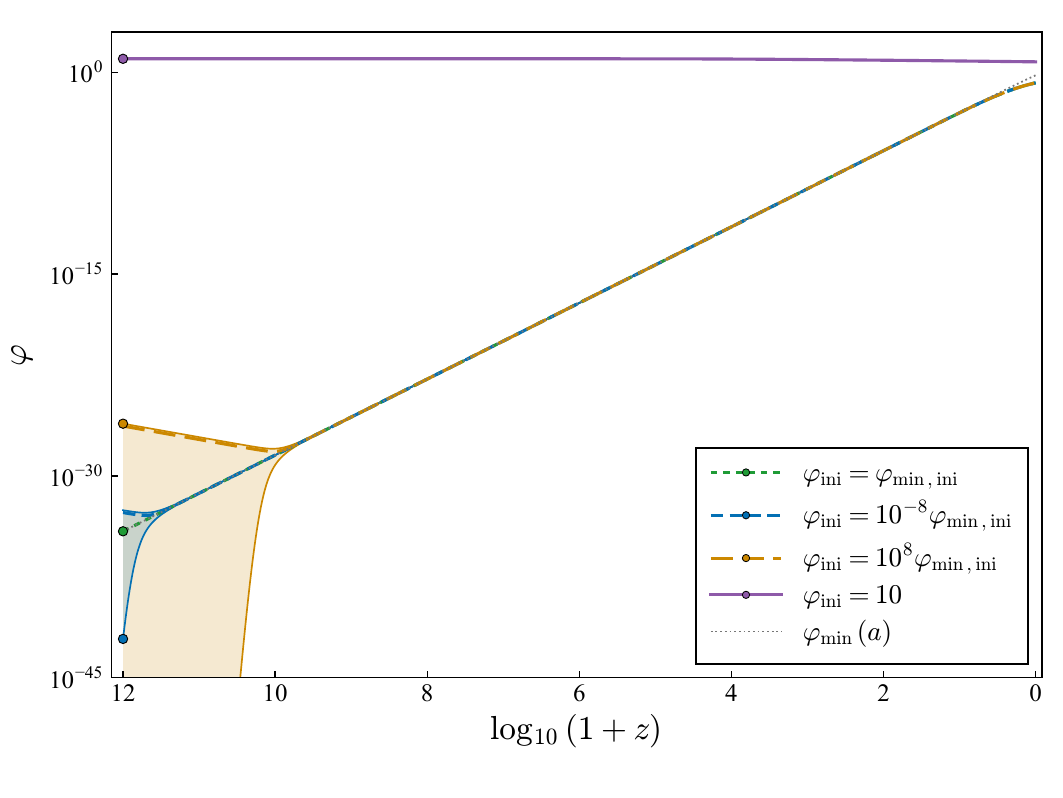}
 \caption{Field evolution for four initial conditions in the $\mathrm{NMCQ}^{+}$ model. Model parameters are selected according to Table~\ref{tab:main}. Here $\varphi_{\min,\mathrm{ini}}\equiv\varphi_{\min}(a_{\mathrm{ini}})$. Dots mark the initial field values. The curves for the three smaller initial field values show cycle averages at early times. The blue and orange shaded regions indicate the oscillation envelopes for the smaller and larger initial field values, respectively. The purple curve corresponds to $\varphi_{\mathrm{ini}}=10$, and the gray dotted curve marks the moving minimum.}
 \label{fig:appendix-initial}
\end{figure}

The stability of $\varphi_{\rm attr}$ as an initial condition has been discussed in Ref.~\cite{Wang:2025znm}.
Here we only examine the response to shifts in the initial field value relative to the minimum.
FIG.~\ref{fig:appendix-initial} compares the trajectories for initial field values $10^{-8}\varphi_{\min}(a_{\mathrm{ini}})$, $10^8\varphi_{\min}(a_{\mathrm{ini}})$, and $10$ with the trajectory initialized at the minimum.
The trajectories averaged over oscillations for the three small initial field values approach the moving minimum and converge to the same slowly evolving solution at late times.
Initial displacements near the minimum induce rapid oscillations, and their amplitude decays through Hubble friction.
However, a sufficiently large initial field value, which may exceed $\varphi_0$, produces a different evolution.
The trajectory with $\varphi_{\mathrm{ini}}=10$ in FIG.~\ref{fig:appendix-initial} is initially almost frozen, then moves toward smaller field values under the coupling force, and remains far from the other three trajectories today.
Along this trajectory, the force from the inverse-power-law bare potential gradient remains weaker than the coupling force.
This difference in the initial conditions may explain why our constraint on the inverse-power-law potential index $n$ is much tighter than that reported in Ref.~\cite{Gomez-Valent:2026ept}.

\begin{figure*}[!t]
\begin{minipage}[t]{\columnwidth}
 \vspace{0pt}
 \centering
 \includegraphics[width=\linewidth]{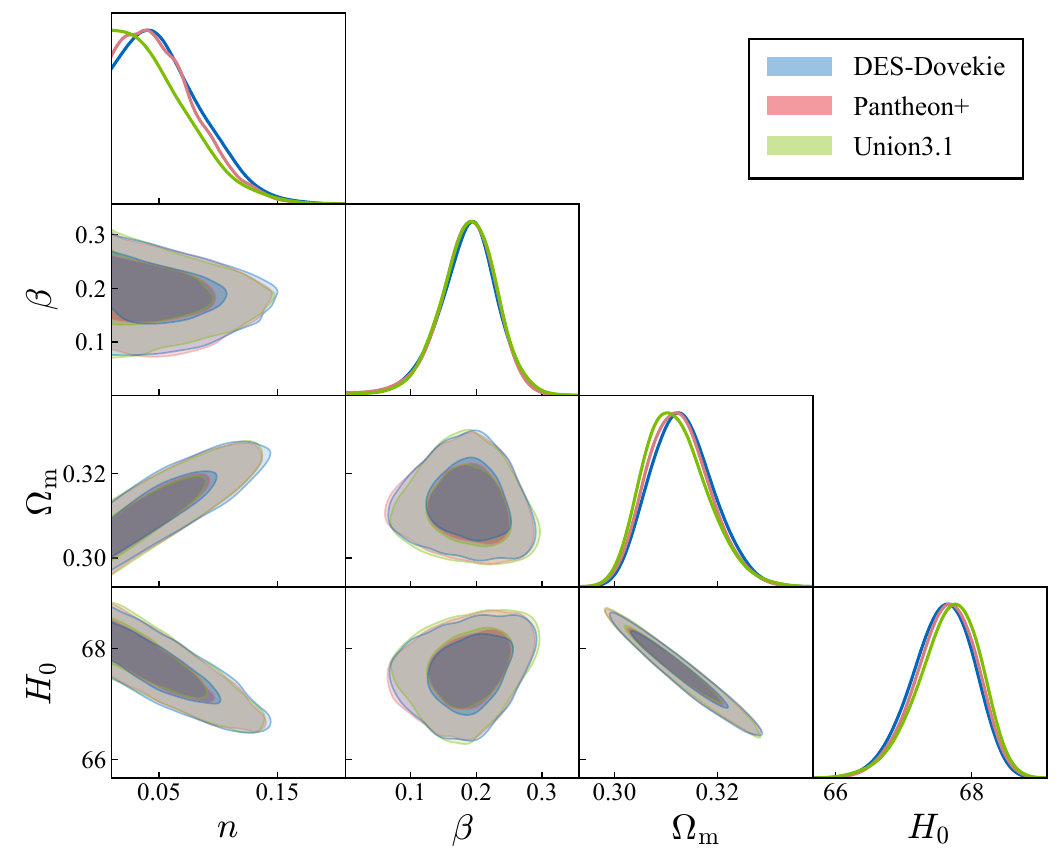}
 \caption{Posteriors of $n$, $\beta$, $\Omega_{\mathrm m}$, and $H_0$ in $\mathrm{NMCQ}^{+}$ for different SN samples. Blue, red, and light green denote DES-Dovekie, Pantheon+, and Union3.1, respectively.}
 \label{fig:appendix-sn}
\end{minipage}\hfill
\begin{minipage}[t]{\columnwidth}
 \vspace{0pt}
 \centering
 \includegraphics[width=\linewidth]{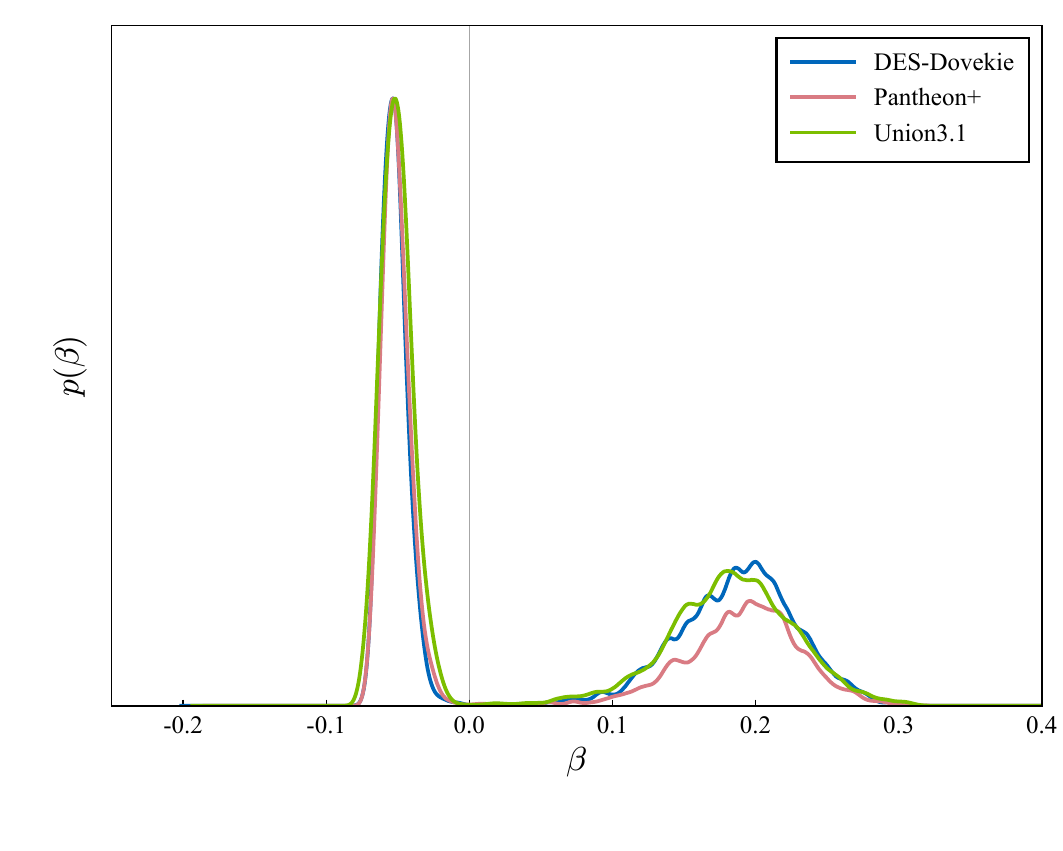}
 \caption{Marginal posteriors of $\beta$ for the three SN samples combined with the same CMB+DESI DR2 BAO data under a common prior $[-0.25,0.4]$.}
 \label{fig:appendix-sn-beta}
\end{minipage}
\end{figure*}

\section{Numerical methods}
\label{sec:appendixB}

The background is normalized to the present scalar field and DM densities.
Spatial flatness fixes the target scalar field density, $\rho_{\varphi0}^{\rm tar}=3H_0^2-\rho_{\mathrm b0}-\rho_{\mathrm{DM},0}-\rho_{\mathrm r0}-\rho_{\nu0}$.
Here $\rho_{\mathrm r0}$ includes photons and massless neutrinos, while $\rho_{\nu0}$ denotes the massive neutrino density.
The inverse-power-law potential amplitude $V_*$ and the trial present-day field value $\varphi_0^*$ of the field today are used as shooting variables, with the DM density written as
\begin{equation}
 \rho_{\mathrm{DM}}(a)=\rho_{\mathrm{DM},0}a^{-3}
 e^{\beta[\varphi(a)-\varphi_0^*]}.
 \label{eq:appendix-dm-normalization}
\end{equation}
The background is integrated to $a=1$, where the shooting conditions are
\begin{equation}
 \rho_\varphi(a=1)=\rho_{\varphi0}^{\rm tar},\qquad
 \varphi(a=1)=\varphi_0^*.
 \label{eq:appendix-shooting}
\end{equation}
These conditions fix the present DE density and ensure that the DM normalization in Eq.~\eqref{eq:appendix-dm-normalization} is consistent with the field evolution.
A damped Newton iteration is used in the logarithmic variables $[\ln(V_*/\rho_{\varphi0}^{\rm tar}),\ln\varphi_0^*]$.
The Jacobian is evaluated by finite differences of the normalized density and field residuals.

Linear scalar perturbations are evolved in synchronous gauge~\cite{Li:2023fdk}.
The field perturbation, DM density contrast, and DM velocity divergence are denoted by $q=\delta\varphi$, $\delta_{\mathrm{DM}}=\delta\rho_{\mathrm{DM}}/\rho_{\mathrm{DM}}$, and $\theta_{\mathrm{DM}}=i\mathbf k\cdot\mathbf v_{\mathrm{DM}}$, respectively.
In terms of $N_{\mathrm{DM}}=\delta_{\mathrm{DM}}-\beta q$, the coupled perturbation equations are
\begin{align}
 N_{\mathrm{DM}}'&=-\theta_{\mathrm{DM}}-\frac{h'}2,
 \label{eq:appendix-number-perturbation}\\
 \theta_{\mathrm{DM}}'&=-(\mathcal H+\beta\varphi')\theta_{\mathrm{DM}}
 +\beta k^2q,
 \label{eq:appendix-velocity-perturbation}
\end{align}
\begin{equation}
 \begin{split}
  &q''+2\mathcal Hq'+\left(k^2+a^2m_{\mathrm{eff}}^2\right)q\\
  &\qquad=-\frac{\varphi'h'}2-\beta a^2\rho_{\mathrm{DM}}N_{\mathrm{DM}},
 \end{split}
 \label{eq:appendix-field-perturbation}
\end{equation}
where $m_{\mathrm{eff}}^2=V_{,\varphi\varphi}+\beta^2\rho_{\mathrm{DM}}$ and $h$ is the trace of the spatial metric perturbation.
The scalar field contributions to the density, pressure, and momentum sources in the Einstein equations are
\begin{equation}
 \begin{aligned}
  \delta\rho_\varphi&=\frac{\varphi'q'}{a^2}+V_{,\varphi}q,\\
  \delta p_\varphi&=\frac{\varphi'q'}{a^2}-V_{,\varphi}q,\\
  (\rho_\varphi+p_\varphi)\theta_\varphi
  &=\frac{k^2\varphi'q}{a^2}.
 \end{aligned}
 \label{eq:appendix-field-sources}
\end{equation}
These sources, together with the perturbations of DM, baryons, photons, and neutrinos, determine the metric evolution.
The photon and neutrino multipole hierarchies follow the standard CAMB treatment.
Adiabatic initial conditions are adopted.
For a heavy field, the initial values of $q$, $q'$, and the metric perturbations are determined jointly from the Hamiltonian constraint and the force balance conditions for the field perturbation.

The large $m_{\mathrm{eff}}$ in $\mathrm{NMCQ}^{+}$ causes severe numerical stiffness.
The scalar perturbation system is therefore evolved implicitly with the fourth-order, L-stable ESDIRK method in SUNDIALS ARKODE.
This six-stage scheme uses a third-order embedded solution to control the step size.
For the linear system $\mathbf Y'=\mathbf J(\eta)\mathbf Y+\mathbf f(\eta)$, each implicit stage satisfies
\begin{equation}
 \begin{split}
  &[\mathbf I-\gamma\Delta\eta\,\mathbf J(\eta_j)]\mathbf Y_j\\
  &\qquad=\mathbf R_j+\gamma\Delta\eta\,\mathbf f(\eta_j),\qquad
  \gamma=\frac14,
 \end{split}
 \label{eq:appendix-implicit-stage}
\end{equation}
where $\mathbf R_j$ contains contributions from the current state and preceding stages.
The stage equations are solved with GMRES, with matrix--vector products evaluated directly from the linear right-hand side.
Step size control accounts for the scales of the perturbation variables and their derivatives, including $|q'|/\Omega_k$ for the field, where $\Omega_k^2=k^2+a^2m_{\mathrm{eff}}^2$.

\section{Supernova sample comparison and full parameter posteriors}
\label{sec:appendixC}

To assess the dependence on the SN sample, DES-Dovekie is replaced by Pantheon+~\cite{Brout:2022vxf} or Union3.1~\cite{Hoyt:2026fve}, with the CMB and DESI DR2 BAO data and priors unchanged.
The likelihoods use relative distances from Pantheon+ and the distance nodes and covariance from the UNITY1.8 analysis of Union3.1.
All SN likelihoods are used without Cepheid distance calibration.
The resulting $\mathrm{NMCQ}^{+}$ parameter constraints are nearly identical, as shown in FIG.~\ref{fig:appendix-sn}.
FIG.~\ref{fig:appendix-sn-beta} compares the marginal posteriors of $\beta$ under the common prior $-0.25\leq\beta\leq0.4$.
All three posteriors have sharp peaks at negative values of $\beta<0$ and broader distributions at positive values $\beta>0$.
The weighted samples give $P(\beta>0)\simeq49.1\%$, $39.5\%$, and $44.0\%$ for DES-Dovekie, Pantheon+, and Union3.1, respectively.
Despite these small differences, neither coupling direction is decisively favored.
Overall, this agreement across the SN samples suggests that the conclusions are unlikely to be driven by statistical fluctuations in a single SN sample, and the conclusions persist across the three tested SN likelihoods.

\begin{figure*}[!t]
 \centering
 \includegraphics[width=\textwidth]{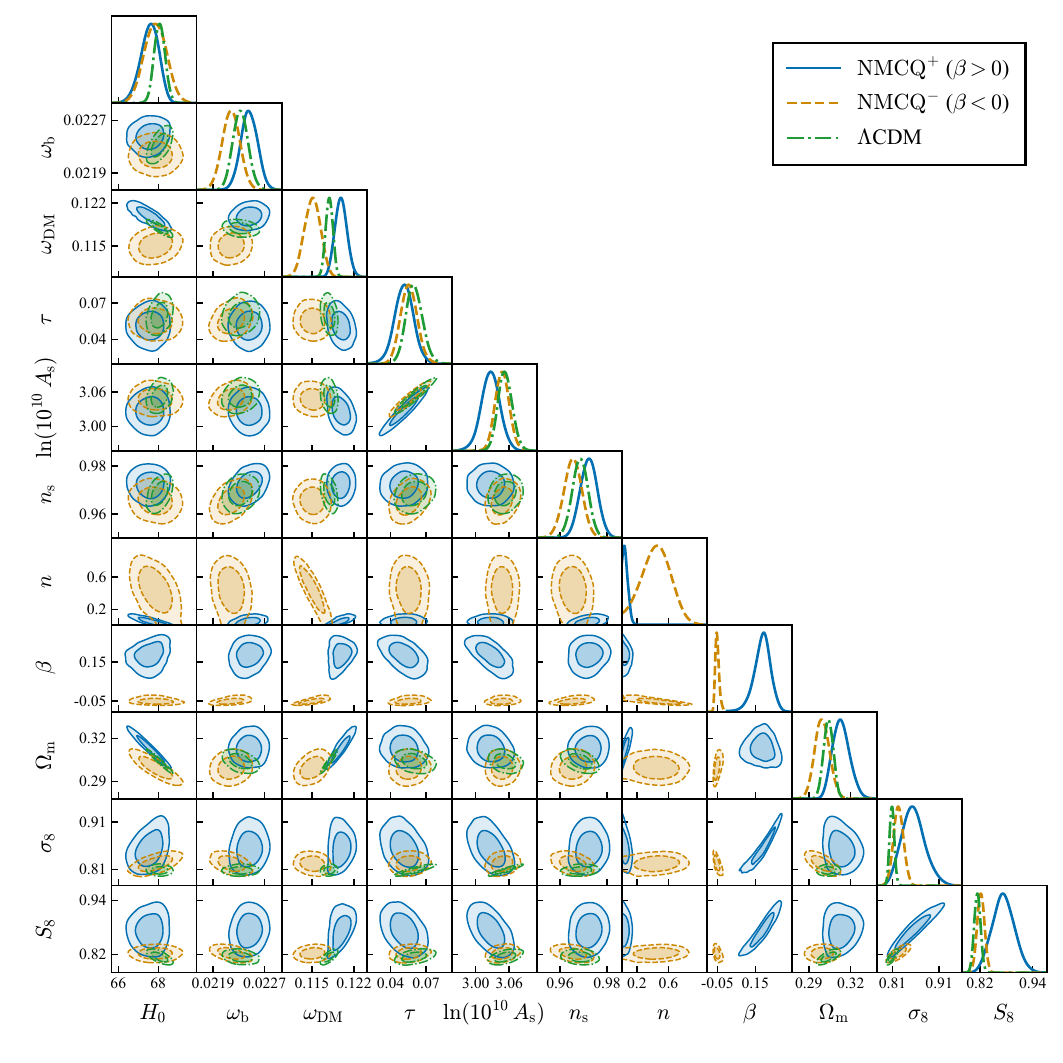}
 \caption{Constraints on cosmological parameters and derived quantities from the baseline CMB+DESI DR2 BAO+DES-Dovekie SN combination. Blue solid, yellow dashed, and green dot-dashed curves correspond to $\mathrm{NMCQ}^{+}$, $\mathrm{NMCQ}^{-}$, and $\Lambda\mathrm{CDM}$, respectively.}
 \label{fig:appendix-main}
\end{figure*}

FIG.~\ref{fig:appendix-main} presents the full cosmological parameter posteriors for the baseline data combination, including the derived distributions of $\Omega_{\mathrm m}$, $\sigma_8$, and $S_8$.
In $\mathrm{NMCQ}^{+}$, $\beta$ is strongly positively correlated with $\sigma_8$ and $S_8$, whereas $n$ and $\beta$ are anticorrelated in $\mathrm{NMCQ}^{-}$.

\bibliography{references}
\end{document}